\documentclass[conference]{IEEEtran}

\usepackage{fancyhdr}
\usepackage{url} 

\usepackage{mathptmx} 
\usepackage[belowskip=-12pt,aboveskip=2pt]{caption}
\usepackage{amsmath}
\usepackage{tikz}
\usepackage{multicol}
\usepackage[flushleft]{threeparttable}
\usepackage{xspace}
\usepackage{multirow}
\usepackage{rotating}
\usepackage{subfig}
\usepackage{chngpage}
\usepackage{amssymb}
\usepackage{pifont}
\usepackage{xcolor}
\usepackage{mathtools}
\usepackage{float}
\usepackage{color}
\usepackage{comment}
\usepackage[ruled,vlined]{algorithm2e}
\usepackage{enumitem}
\usepackage{amsfonts}
\usepackage{listings}
\usepackage{tabularx}
\PassOptionsToPackage{hyphens}{url}
\usepackage{hhline}
\usepackage{nccmath}
\usepackage[normalem]{ulem}
\usepackage{soul} 
\usepackage{cite}

\def\BibTeX{{\rm B\kern-.05em{\sc i\kern-.025em b}\kern-.08em
   T\kern-.1667em\lower.7ex\hbox{E}\kern-.125emX}}

\newcommand{\bheading}[1]{{\vspace{2pt}\noindent{\textbf{#1}}\hspace{2pt}}}

\newcommand{\cut}[1]{}

\def\authnotes{1}
\newcommand{\authnote}[2]{\ifnum\authnotes=1\begin{quote}\textbf{#1 says:} #2\end{quote}\fi}
\newcommand{\fixme}[1]{\ifnum\authnotes=1\textbf{\textcolor{red}{[FIXME: #1]}}\fi}

\newcommand{\gyreffig}[1]{Fig.~\ref{#1}}
\newcommand{\gyrefsec}[1]{Section~\ref{#1}}
\newcommand{\gyreftbl}[1]{Table~\ref{#1}}

\newcommand{\opsetup}{\emph{Setup}}
\newcommand{\opauth}{\emph{Authorize}}
\newcommand{\opaccess}{\emph{Access}}
\newcommand{\opuse}{\emph{Use}}
\newcommand{\opsend}{\emph{Send}}
\newcommand{\oprecv}{\emph{Receive}}
\newcommand{\defsetup}{\textit{No Setup}}
\newcommand{\defaccess}{\textit{No Access without Authorization}}
\newcommand{\defuse}{\textit{No Use without Authorization}}
\newcommand{\defsend}{\textit{No Send without Authorization}}
\newcommand{\defnospec}{\textit{No Speculation}}
\newcommand{\fullspec}{\textit{Unconstrained Speculation}}

\definecolor{dkgreen}{rgb}{0,0.6,0}
\definecolor{gray}{rgb}{0.5,0.5,0.5}
\definecolor{mauve}{rgb}{0.58,0,0.82}

\begin{document}

\title{SoK: Hardware Defenses Against Speculative Execution Attacks}

\author{\IEEEauthorblockN{Guangyuan Hu}
\IEEEauthorblockA{\textit{Princeton University}\\
gh9@princeton.edu}
\and
\IEEEauthorblockN{Zecheng He}
\IEEEauthorblockA{\textit{Princeton University}\\
zechengh@princeton.edu}
\and
\IEEEauthorblockN{Ruby B. Lee}
\IEEEauthorblockA{\textit{Princeton University}\\
rblee@princeton.edu}
}

\maketitle

\begin{abstract}

Speculative execution attacks leverage the speculative and out-of-order execution features in modern computer processors to access secret data or execute code that should not be executed. Secret information can then be leaked through a covert channel. While software patches can be installed for mitigation on existing hardware, these solutions can incur big performance overhead. Hardware mitigation is being studied extensively by the computer architecture community. It has the benefit of preserving software compatibility and the potential for much smaller performance overhead than software solutions.

This paper presents a systematization of the hardware defenses against speculative execution attacks that have been proposed. We show that speculative execution attacks consist of 6 critical attack steps. We propose defense strategies, each of which prevents a critical attack step from happening, thus preventing the attack from succeeding. We then summarize 20 hardware defenses and overhead-reducing features that have been proposed. We show that each defense proposed can be classified under one of our defense strategies, which also explains why it can thwart the attack from succeeding. We discuss the scope of the defenses, their performance overhead, and the security-performance trade-offs that can be made.

\end{abstract}

\pagestyle{plain}

\section{Introduction} \label{sec_intro}

Speculative execution attacks, also known as transient execution attacks, are a serious security problem. They exploit performance enhancement features in hardware to access secret data and leak this secret out through microarchitectural covert channels. This negates the confidentiality and integrity protections provided by software isolation, and also by hardware isolation features such as secure enclaves \cite{SGX, TEE}.

In particular, Spectre \cite{spectre}, Meltdown \cite{meltdown} and Foreshadow \cite{van2018foreshadow} bypass the isolation across processes and privilege levels. The Spectre attack bypasses the memory protection provided by software bounds checking, while the Meltdown attack breaches the memory isolation between the kernel and a user application. Foreshadow \cite{van2018foreshadow}, and its variants Foreshadow-OS and Foreshadow-VMM \cite{weisse2018foreshadowNg}, breach the Intel SGX enclave isolation, user-to-kernel memory isolation, and virtual-machine-to-hypervisor isolation, respectively.

The severity of these attacks has resulted in many specific fixes for specific attack variants implemented by the computer industry. These include using instructions to serialize execution \cite{IntelSepcAttack, ARMSpectre}, to flush hardware prediction states \cite{IBRS}, to avoid using untrusted predictions \cite{Retpoline}, and to restrict accesses to secret information \cite{JavaScriptSpectre, Webkitdefense, Chromedefense}. However, most of these solutions require changes to the existing software. Furthermore, they also cause significant performance overhead (at least 2X slower \cite{JavaScriptSpectre}, sometimes up to 8X). Last but not least, the software countermeasures are usually attack-specific. New patches are required to effectively protect against the emerging attacks, which is neither efficient nor sustainable.

In response to these attacks on hardware microarchitecture performance optimization features, there have been proposals of hardware defenses as well as features that reduce the performance overhead of defenses \cite{invisispec, dawg, condspec, csf, spectreguard, safespec, efficientspec, specshield, stt, nda, cleanupspec, mi6, ironhide, context, SamsungExynosCPU, muontrap, sdo, clearshadow, invarspec, dolma}, which we show in \gyreftbl{tbl_defenses} in chronological order. One key advantage is that the hardware solution can monitor the instruction execution status and accurately protect against speculative vulnerabilities. Another advantage of some hardware solutions is their non-intrusive interaction with the existing software, while inducing low performance overhead. These hardware defenses can read the unmodified program but delay or change the execution of secret-leaking instructions so that the information leakage through hardware states is eliminated. Some microarchitectural defenses also allow security-performance trade-offs and overhead-reducing features \cite{sdo, clearshadow, invarspec}. 

\begin{table}[t]
\resizebox{\columnwidth}{!}{
\begin{tabular}{|l|l|l|}
\hline
\textbf{Defense and Overhead-reducing Feature}                    & \textbf{Conference}                                       & \textbf{Year} \\ \hline
InvisiSpec \cite{invisispec}                          & MICRO                                                     & 2018          \\ \hline
DAWG \cite{dawg}                                      & MICRO                                                     & 2018          \\ \hline
CondSpec \cite{condspec}                              & HPCA                                                      & 2019          \\ \hline
Context-sensitive fencing (CSF) \cite{csf}            & ASPLOS                                                    & 2019          \\ \hline
SpectreGuard \cite{spectreguard}                      & DAC                                                       & 2019          \\ \hline
SafeSpec \cite{safespec}                              & DAC                                                       & 2019          \\ \hline
EfficientSpec \cite{efficientspec}                    & ISCA                                                      & 2019          \\ \hline
SpecShield \cite{specshield}                          & PACT                                                      & 2019          \\ \hline
STT \cite{stt}                                        & MICRO                                                     & 2019          \\ \hline
NDA \cite{nda}                                        & MICRO                                                     & 2019          \\ \hline
CleanupSpec \cite{cleanupspec}                        & MICRO                                                     & 2019          \\ \hline
MI6 \cite{mi6}                                        & MICRO                                                     & 2019          \\ \hline
IRONHIDE \cite{ironhide}                              & HPCA                                                      & 2020          \\ \hline
ConTExT \cite{context}                                & NDSS                                                      & 2020          \\ \hline
Predictor state encryption \cite{SamsungExynosCPU}    & ISCA                                                      & 2020          \\ \hline
MuonTrap \cite{muontrap}                              & ISCA                                                      & 2020          \\ \hline
Speculative Data-Oblivious Execution (SDO) \cite{sdo} & ISCA                                                      & 2020          \\ \hline
Clearing the Shadows \cite{clearshadow           }& PACT                                                      & 2020          \\ \hline
InvarSpec \cite{invarspec                            }& MICRO                                                     & 2020          \\ \hline
DOLMA \cite{dolma}                                    & \begin{tabular}[c]{@{}l@{}}USENIX\\ Security\end{tabular} & 2021          \\ \hline
\end{tabular}
}
\caption{Hardware defenses and overhead-reducing features against speculative execution attacks published in recent computer architecture and security conferences.}
\label{tbl_defenses}
\end{table}

However, the working mechanisms and scope of different hardware defenses have not been systematically described and compared. Hence, our goal in this paper is to systematize the hardware defenses, to illustrate their key similarities and differences, and to assist future researchers to more easily understand and reason about how an existing defense works. While the goal of this paper is not to describe all the speculative execution attacks in detail, as there are many past work surveying and summarizing these \cite{AttackSummaryMS, AttackSummaryGraz:236214, AttackSummaryYale:2005.13435, AttackSummaryPALMSHPCA}, we analyze the critical attack steps of 23 speculative execution attacks. We then show how the hardware defenses mitigate the attacks by preventing these steps, connecting the attacks and defenses.


Our key contributions are:
\begin{itemize}
\item Producing attack taxonomies based on secret access or secret leakage, covering 23 variants of speculative execution attacks. 
\item Defining new defense strategies based on preventing at least one of the critical attack steps. 
\item Producing a new taxonomy of 4 hardware defense strategies and lower-level categories of defenses. 
\item Creating a systematized view and description of 20 representative hardware defenses and overhead-reducing features proposed to date. 
\item Presenting the performance overhead of the defenses, and illustrating security-performance tradeoffs. 
\end{itemize}

\section{Microarchitecture and Covert Channel Background}
\label{sec_background}

We first describe hardware performance optimization features that can be exploited for speculative execution attacks. 

\bheading{Out-of-Order (OoO) execution.}~An Out-of-Order (OoO) processor
is a microarchitecture performance enhancement feature used to boost the throughput of processors by allowing instructions later in the program order to execute before the previous instructions have completed. For example, an earlier instruction may be waiting for one of its operands, or for a functional unit or memory to free up, or for determining if a branch should be taken or where to branch to. Later instructions in an in-order processor that have no dependencies will have to wait unnecessarily. In contrast, an Out-of-Order processor allows the instructions with no dependencies to execute immediately, as long as they retire in-order.

\begin{figure*}[t]
    \centering
    \includegraphics[width=0.85\linewidth]{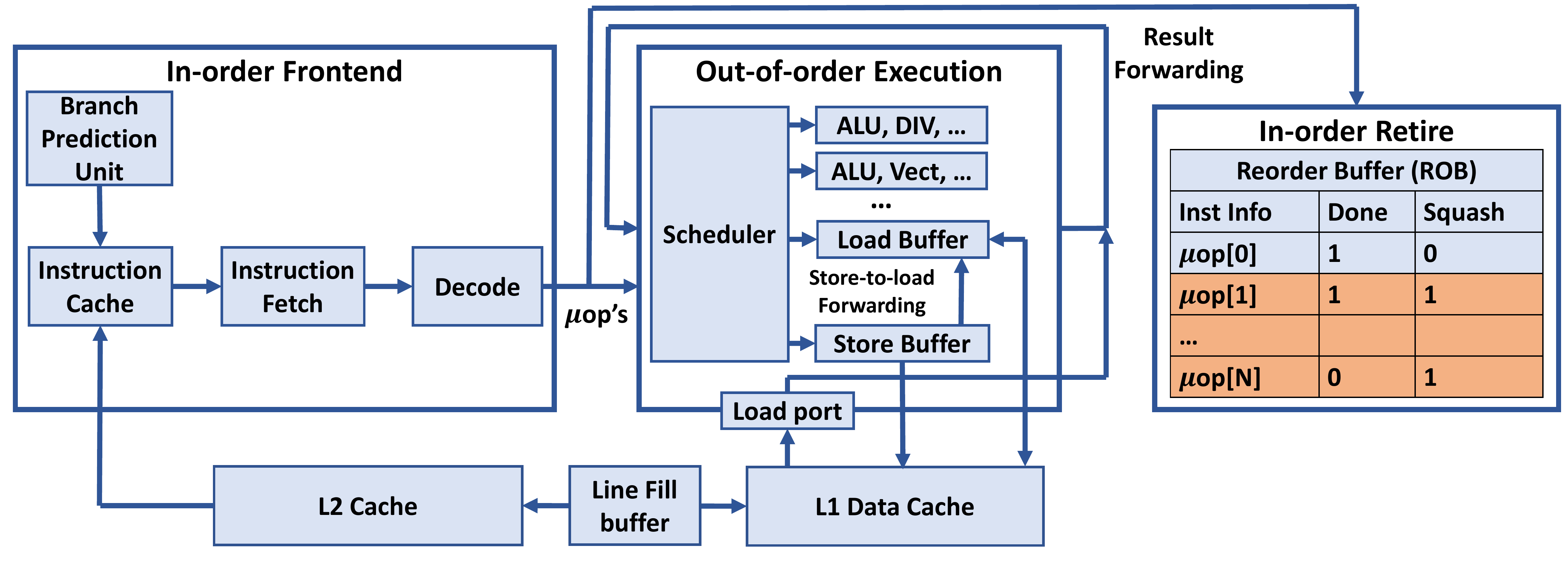}
    \caption{\small{A block diagram of a typical out-of-order processor. The contents of the ROB show a sample situation where instruction $\mu$op[0] executed correctly, but $\mu$op[1] to $\mu$op[N] were executed speculatively and incorrectly and had to be squased.}}
    \label{img_squash}
\end{figure*}

\gyreffig{img_squash} shows a generic Out-of-Order (OoO) processor where instructions are fetched in program order but executed Out-of-Order. Instructions are forced to retire in-order to maintain precise exceptions, i.e., if an instruction results in an exception, the following instructions must be ``squashed'' as if they were never executed. We will use this generic OoO model to explain the defenses in a unified way in the rest of the paper.

Instructions are fetched and decoded to
microarchitecture-level operations (denoted $\mu$op's) in program order, but after the $\mu$op's are dispatched to the execution stage, the hardware scheduler can schedule any ready $\mu$op's to different functional units for execution. Thus, the execution of later $\mu$op's can complete earlier than those of previous instructions, which also allows the results of these $\mu$op's to be used earlier.
The result from the execution of a $\mu$op is \emph{forwarded} and used by other dependent $\mu$op's. 

An important microarchitecture structure, that we will refer to in discussing hardware defenses, is the \emph{Re-Order Buffer (ROB)} shown in \gyreffig{img_squash}. The ROB records the instruction's or $\mu$op's information as well as its execution status, such as whether the instruction has finished its execution (\emph{Done} = ``1'' in the figure) and whether the instruction should be squashed (\emph{Squash} = ``1''). The ROB guarantees that if an instruction needs to be squashed, all the subsequent instructions are also squashed. The ROB also acts as a FIFO queue to preserve the program order so an instruction can only retire when it reaches the head of the ROB.

\bheading{Speculative execution.}
Speculative execution
is a further performance enhancement
that allows instructions to be tentatively (i.e., speculatively) executed, even when the control flow has not been determined,
or the data from memory has not arrived. For instance, when the processor fetches a branch whose operand is not available, e.g., having to be read from memory, the address of the next instruction is predicted and fetched so that the processor does not have to stall its pipeline. If the prediction is found to be correct later, the speculative execution improves the performance by executing code on the correct path in advance. However, if the prediction is found to be incorrect, the processor needs to flush the pipeline so that the results of the speculatively executed instructions are discarded. This is called a \textit{squash}, where the processor restores the architectural state, e.g., the register values visible to the software, as if the mispredicted instructions have not executed.

\bheading{Hardware predictors.}~From the microarchitecture perspective, speculative execution happens because hardware predictors are present that allow tentative forward progress even when an instruction has unresolved dependencies. For conditional branch instructions, branch predictors predict whether the branch will be taken or not.  For indirect branches, the Branch Target Buffer (BTB) predicts the target address. For return instructions, the Return Stack Buffer (RSB) or Return Address Stack (RAS) predicts the address to return to after a procedure call. 

\bheading{Microarchitectural state and covert channels.}
Microarchitectural states are the states of hardware units that are not directly accessible to the programmer or software. Even if invisible from the software's view, these states can impact the execution time of certain programs and the states can be inferred if the processor executes these programs. If one program modifies a certain microarchitectural state with another monitoring it, these two programs form a microarchitectural covert channel in which the former is the sender and the latter is the receiver. Examples include the addresses of cache lines in various cache levels, which we describe in detail below, and the busy status of different hardware resources. 

\bheading{Cache state and covert channel.}
One critical microarchitectural state is the cache state. A cache has many cache lines corresponding to different addresses. Since cache hits are fast, and cache misses are slow, cache timing attacks are possible, leaking information through observing the cache access time. 

One example exploiting a cache covert channel is the \textit{flush-reload} technique \cite{flushreload:184415}, where the sender is an insider and the receiver an outsider attacker. 
During the setup phase of the covert channel, certain cache lines are flushed out of the cache. To send a secret out, the sender accesses a secret-dependent address, which brings back one of the flushed lines. 
The receiver will later measure the time to reload each cache line and infer whether this cache line is fetched by the sender by observing whether it is a cache hit. The flush-reload cache covert channel is used in most of the speculative execution attacks published. 

There are other techniques for covert communication through cache state. In a \textit{prime-probe} \cite{primeprobe} covert channel, the receiver first primes the cache to fill the cache with its own cache lines. The sender then accesses certain addresses, evicting some of the receiver's cache lines. The receiver can get to know which cache lines the sender accessed by loading each cache line and observing cache misses. In a \textit{flush-flush} \cite{flushflush} covert channel, the receiver keeps evicting certain addresses by executing the flush instruction. If the sender accesses some of these addresses and brings them into the cache, the time to flush will be longer, so the receiver can infer information from timing the second flush.

Many other types of covert or side channels, not using caches, nor timing, are also possible.
\section{Speculative Execution Attacks}
\label{sec_attack}

We first present some critical attack steps that we have identified in existing speculative execution attacks.

\subsection{Critical Attack Steps}
\label{sec_attack_char}

\begin{figure}[t]
    \centering
    \includegraphics[width=\linewidth]{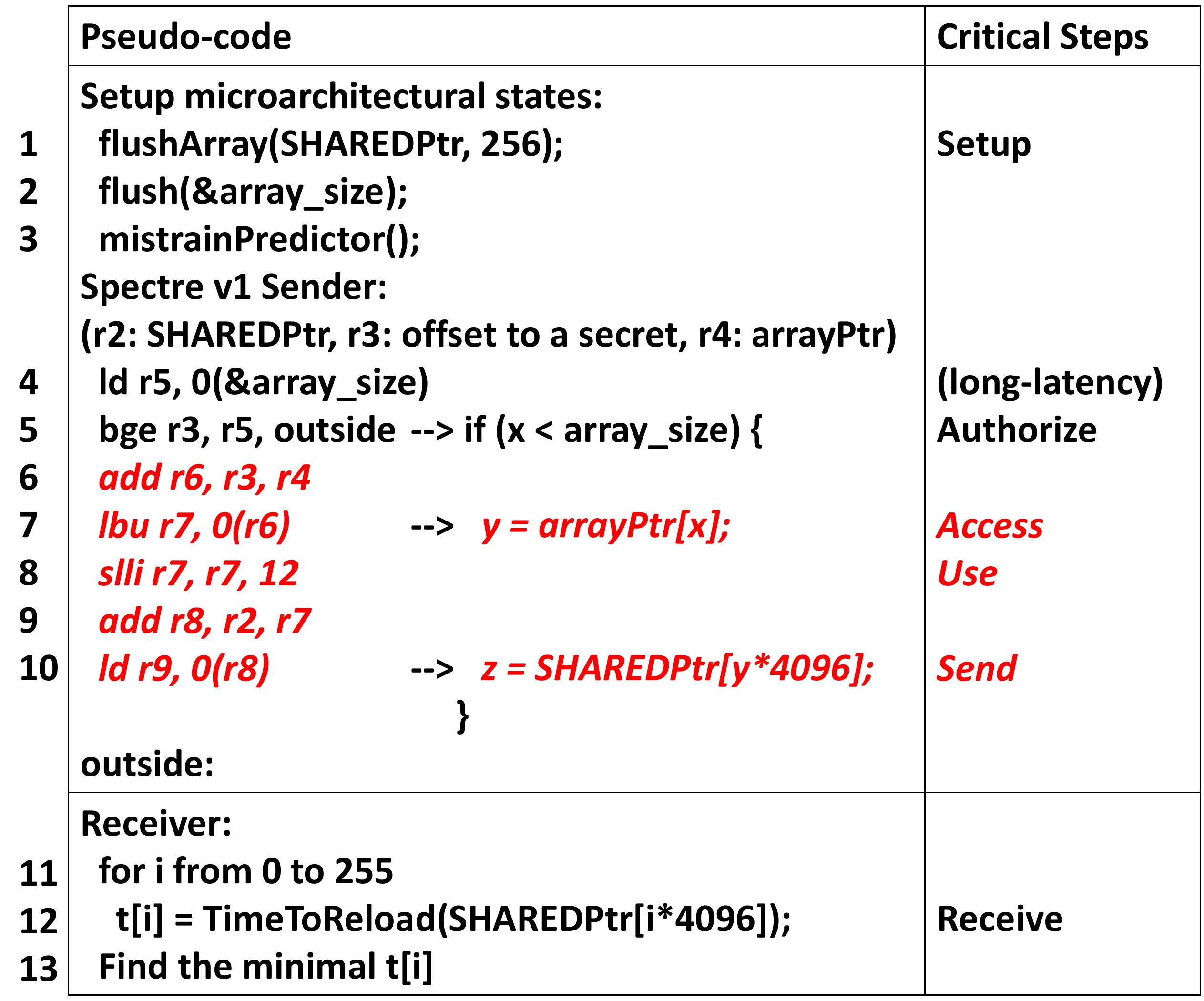}
    \caption{\small{Spectre v1 attack (bypassing array bounds checking). The assembly code and pseudo code of the attack bypass control flow authorization by a conditional branch to access a secret. The attack leaks an 8-bit secret through the most commonly used flush-reload cache covert channel by loading in a cache line in the shared array into the cache. The code in red is the transient execution that will be squashed. The comments after the arrows show the high-level language equivalents of the assembly code.}}
    \label{img_code_spectre}
\end{figure}

Although the exact workflow of an attack may vary, we observe that they all consist of 6 critical steps. These are shown in the right column of \gyreffig{img_code_spectre} and described below.

\bheading{Setup.} The \emph{Setup} step sets up the initial hardware state, e.g., the branch predictor state for Spectre v1, so that the processor will enter speculative execution. It also sets up the initial state for the covert channel, e.g., flushing the shared cache lines for a flush-reload channel.

\bheading{Authorize.} The attack starts with the \emph{Authorize} step. The \emph{Authorize} operation performs the authorization required for accessing a memory location or a protected register. For speculative attacks, the speculative execution window starts when the authorization is delayed.

\bheading{Access.} When the authorization is delayed, the \emph{Access} step in a speculative attack can read a secret from the cache, the memory, a protected register or a microarchitectural buffer that is otherwise not allowed.

\bheading{Use.} The \emph{Use} step uses the secret to generate a secret-dependent operation. Examples are instructions that compute a memory address for a later load operation.

\bheading{Send.} The \emph{Send} step alters the microarchitectural state of the covert channel in a secret-dependent way. Even if the \emph{access}, \emph{use} and \emph{send} operations will all be squashed after the authorization fails, the microarchitectural state change may remain and can be discovered later by the receiver.

\bheading{Receive.} The recovery of the secret from the covert channel by the attacker.

\subsection{A Spectre v1 Attack Example}

For concreteness, let us first consider a particular speculative attack, the Spectre v1 attack. In \gyreffig{img_code_spectre}, we show the pseudo-code of the Spectre v1 attack and the RISC assembly instructions executed during speculative execution. Lines 1-3 \emph{set up} the microarchitectural state. The cache lines containing the shared array pointed by SHAREDPtr are flushed from caches as the preparation for the flush-reload cache covert channel which we described in \gyrefsec{sec_background}. The size of the private array pointed to by arrayPtr is also flushed so that the load on line 4 will take a long time to finish. Also, the branch predictor is mistrained so that the prediction of the conditional branch in line 5 will be ``not taken''. The conditional branch, bge, in line 5 performs the \emph{authorization} for the later load byte instruction, lbu, which \emph{accesses} the secret byte in line 7.  Since the conditional branch checking is delayed by the previous load instruction in line 4, a branch predictor is invoked.  Due to the mistraining, the branch is not taken and the secret is illegally accessed by the lbu instruction. In line 8 and 9, the secret is then \emph{used} to calculate a memory address of the next ld instruction in line 10. This ld instruction is a covert \emph{send} instruction that leaks out the secret through the cache covert channel. In line 11-13, the receiver measures the latency to access the shared array to find out which memory address in the shared array was accessed by the sender. The memory address that hits in the cache leaks the secret.

\subsection{Other Attacks}

\gyreftbl{img_attacks_table} gives a listing of the speculative attacks published to date \cite{spectre, kiriansky2018speculative, schwarz2019netspectre, spectrersb, ret2spec:10.1145/3243734.3243761, spectressb, spectrev3a, lazyfp, meltdown, van2018foreshadow, weisse2018foreshadowNg, van2019ridl, intelMDS, schwarz2019zombieload, canella2019fallout, TAA, VRS, cacheout, intelL1DES, crosstalk, intelSRBDS, vanbulck2020lvi, speculativeinterference}. We show their Common Vulnerabilities and Exposures (CVE) numbers, description and publication date.
All the attack variants in \gyreftbl{img_attacks_table}, except for the last speculative interference attack, introduce a new way to bypass authorization to access the secret. The speculative interference attack introduces a new way to change the timing of non-speculative instructions, which adds a new dimension to the covert \opsend~operation.


\begin{table}[t]
\resizebox{\columnwidth}{!}{
\begin{tabular}{|l|l|l|l|}
\hline
\textbf{Attack}                                                                                                                       & \textbf{CVE}   & \textbf{Description}                                                                                & \textbf{Date} \\ \hline
Spectre v1 \cite{spectre}                                                                                            & 2017-5753  & \begin{tabular}[c]{@{}l@{}}Speculative boundary check\\ bypass for read\end{tabular}                & 2018.1        \\ \hline
Spectre v1.1 \cite{kiriansky2018speculative}                                                                         & 2018-3693  & \begin{tabular}[c]{@{}l@{}}Speculative boundary check\\ bypass for write\end{tabular}               & 2018.7        \\ \hline
NetSpectre \cite{schwarz2019netspectre}                                                                              & 2017-5753  & \begin{tabular}[c]{@{}l@{}}Remote attack performing a\\ bounds check bypass\end{tabular}            & 2018.1        \\ \hline
Spectre v2 \cite{spectre}                                                                                            & 2017-5715  & Branch target misprediction                                                                         & 2018.1        \\ \hline
Spectre RSB\cite{spectrersb, ret2spec:10.1145/3243734.3243761}                                                       & 2018-15572 & Return target misprediction                                                                         & 2018.8        \\ \hline
Spectre SSB \cite{spectressb}                                                                                        & 2018-3639  & \begin{tabular}[c]{@{}l@{}}Speculative store bypass, read \\ stale data in memory\end{tabular}      & 2018.5        \\ \hline
\begin{tabular}[c]{@{}l@{}}Meltdown-Reg\\ (Spectre v3a) \cite{spectrev3a}\end{tabular}                               & 2018-3640  & \begin{tabular}[c]{@{}l@{}}System register value leakage\\ to unprivileged attacker\end{tabular}    & 2018.5        \\ \hline
Lazy FP \cite{lazyfp}                                                                                                & 2018-3665  & Leak of FPU state                                                                                   & 2018.6        \\ \hline
\begin{tabular}[c]{@{}l@{}}Meltdown \\ (Spectre v3) \cite{meltdown}\end{tabular}                                     & 2017-5754  & \begin{tabular}[c]{@{}l@{}}Kernel content leakage to\\ unprivileged attacker\end{tabular}           & 2018.1        \\ \hline
\begin{tabular}[c]{@{}l@{}}Foreshadow (L1 \\ Terminal Fault) \cite{van2018foreshadow}\end{tabular}                   & 2018-3615  & SGX enclave memory leakage                                                                          & 2018.8        \\ \hline
Foreshadow-OS \cite{weisse2018foreshadowNg}                                                                          & 2018-3620  & OS memory leakage                                                                                   & 2018.8        \\ \hline
Foreshadow-VMM \cite{weisse2018foreshadowNg}                                                                         & 2018-3646  & VMM memory leakage                                                                                  & 2018.8        \\ \hline
Spectre v1.2 \cite{kiriansky2018speculative}                                                                         & N/A            & \begin{tabular}[c]{@{}l@{}}Speculative write to \\ read-only memory\end{tabular}                    & 2018.7        \\ \hline
RIDL/MLPDS \cite{van2019ridl, intelMDS}                                                                              & 2018-12127 & MDS leakage from load port                                                                          & 2019.5        \\ \hline
\begin{tabular}[c]{@{}l@{}}RIDL/ZombieLoad/\\ MFBDS \cite{van2019ridl, schwarz2019zombieload, intelMDS}\end{tabular} & 2018-12130 & MDS leakage from line fill buffer                                                                   & 2019.5        \\ \hline
Fallout/MSBDS \cite{canella2019fallout, intelMDS}                                                                    & 2018-12126 & MDS leakage from store buffer                                                                       & 2019.5        \\ \hline
TAA \cite{TAA}                                                                                                       & 2019-11135 & TSX Asynchronous Abort                                                                              & 2019.11       \\ \hline
RIDL/MDSUM \cite{van2019ridl, intelMDS}                                                                              & 2019-11091 & \begin{tabular}[c]{@{}l@{}}MDS leakage from \\ uncacheable memory\end{tabular}                      & 2019.5        \\ \hline
VRS \cite{VRS}                                                                                                       & 2020-0548  & Vector Register Sampling                                                                            & 2020.1        \\ \hline
CacheOut/L1DES\cite{cacheout, intelL1DES}                                                                            & 2020-0549  & L1D Eviction Sampling                                                                               & 2020.1        \\ \hline
\begin{tabular}[c]{@{}l@{}}CROSSTALK/\\ SRBDS \cite{crosstalk, intelSRBDS}\end{tabular}                              & 2020-0543  & \begin{tabular}[c]{@{}l@{}}Special Register Buffer\\ Data Sampling\end{tabular}                     & 2020.6        \\ \hline
LVI \cite{vanbulck2020lvi}                                                                                           & 2020-0551  & \begin{tabular}[c]{@{}l@{}}Load Value Injection causing\\ memory disclosure\end{tabular}            & 2020.3        \\ \hline
\begin{tabular}[c]{@{}l@{}}Speculative\\ Interference\cite{speculativeinterference}\end{tabular}                     & N/A            & \begin{tabular}[c]{@{}l@{}}Speculative interference on non-\\ speculative instructions\end{tabular} & 2020.9        \\ \hline
\end{tabular}
}
    \caption{\small{The Speculative (Transient) Execution Attack variants. \textit{Date} is year.month of publication.}}
    \label{img_attacks_table}
\end{table}

\begin{figure*}[t]
    \centering
    \includegraphics[width=\linewidth]{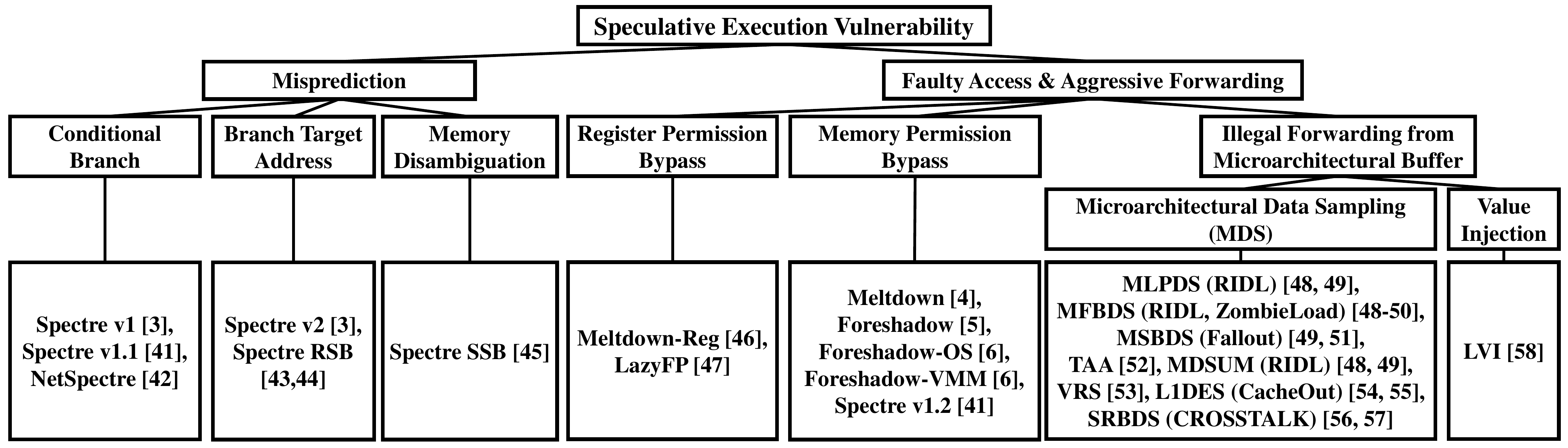}
    \caption{\small{Taxonomy of secret access (bypassed authorization and secret access steps). The third and fourth rows show the hardware mechanisms used to trigger the transient execution.  They correspond to delayed \emph{Authorize} operations that are temporarily bypassed. The last row shows the attacks that exploit these hardware features. These are listed in the same order as in Table \ref{tbl_defenses}, from left to right.}}
    \label{img_attacks_characterization}
\end{figure*}

\bheading{Hardware features for malicious speculative execution.} In \gyreffig{img_attacks_characterization}, we show the hardware features that can be exploited to launch malicious speculative execution attacks, especially to access a secret. 

The first major category of features causing misprediction include the conditional branch prediction, the prediction for branch target address and the memory disambiguation. Spectre v1 \cite{spectre} attack mistrains the conditional branch for bounds checking to read an out-of-bounds secret. Spectre v1.1 \cite{kiriansky2018speculative} also uses misprediction for conditional branch to bypass bounds checking but performs an out-of-bounds write during speculative execution. Even if the write to memory will not become visible, the write may change a jump target, e.g., the return address, and execute an \opaccess-\opuse-\opsend~gadget (i.e., a code snippet) as we show in lines 7-10 in \gyreffig{img_code_spectre} to read and leak a secret. NetSpectre \cite{schwarz2019netspectre} shows that the mistraining of the conditional branch predictor can be performed remotely.

Another control-flow misprediction based attack is the Spectre v2 attack \cite{spectre}, which injects a malicious target into the branch target buffer (BTB) for indirect branches. Similarly, the Spectre RSB attack \cite{spectrersb, ret2spec:10.1145/3243734.3243761} injects wrong return addresses into the return stack buffer (RSB) for function returns. Both can cause information leakage by directing the control flow to an \opaccess-\opuse-\opsend~gadget.

Memory disambiguation checks whether the value written by a previous store instruction, which has not yet been written back to the cache-memory system, should be forwarded to a later load instruction that reads from the same address. In the Speculative Store Bypass (Spectre SSB) attack, if the store address has not been computed and the processor predicts that the addresses of the current load and a previous store are different, then stale data, which can be a secret, can be loaded from the memory system to the processor and get leaked out.

The second major category of hardware features exploited consists of an illegal access that reads a secret and forwards it to dependent instructions before it is squashed. We call these "faulty access and aggressive forwarding" attacks. The first type of attacks transiently bypasses permission checks of special registers and delays the exception handling. Meltdown-Reg \cite{spectrev3a} can read the system parameter stored in a system register while LazyFP \cite{lazyfp} leaks the stale floating-point unit (FPU) state of a previous domain that is not cleared until first used in a new context.

The second type of faulty access attacks transiently violate memory access permission checking and reads illegal data with a memory access instruction. Meltdown \cite{meltdown} reads and leaks kernel data before the execution is squashed due to the failed supervisor permission check of the secret access. The Foreshadow (L1 terminal fault) attack variants \cite{van2018foreshadow, weisse2018foreshadowNg} exploit loads which do not have a valid virtual address to physical address mapping. The address translation will abort prematurely by returning a partially translated address. If a secret at this incorrect address is present in the L1 cache, it can be speculatively accessed and leaked out. The leaked data can be a secret in an SGX enclave (Foreshadow), in the kernel space (Foreshadow-OS) or in the virtual machine monitor space (Foreshadow-VMM). Spectre v1.2 attack \cite{kiriansky2018speculative} transiently bypasses the read/write permission and writes to a read-only address. The illegal write can trigger an \opaccess-\opuse-\opsend~gadget to leak a secret if it is a branch target.

The more recent type of attacks (in 2019 and 2020) exploit the hardware vulnerability that some stale data, which is stored in microarchitectural buffers can be  read by a load that will cause a fault or invoke a microcode assist \cite{intelMDS}. The data can belong to another security domain and can be at a different address from the address the faulting load is accessing. This type of attack is called a microarchitectural data sampling (MDS) attack. In an MDS attack, the victim program first executes and accesses a secret. The secret can be temporarily stored in a microarchitectural buffer when it is in-flight. However, the stale secret value can be forwarded to a faulting or microcode-assisted load issued by the MDS attacker which then sends it out through a covert channel.

Microarchitectural buffers that have been shown to store stale secret values include the load port, the line fill buffer and the store buffer, which we show in \gyreffig{img_squash}. The load port temporarily stores the data when it is read by a load operation and being written into a register. The line fill buffer stores a memory line that missed in the L1 data cache and is being returned from the L2 cache \cite{intelMDS}. The store buffer stores the data and addresses of store operations to be written to the L1 data cache. RIDL \cite{van2019ridl} leaks the secret stored in the load port called Microarchitectural Load Port Data Sampling (MLPDS) \cite{intelMDS} and the line fill buffer called Microarchitectural Fill Buffer Data Sampling (MFBDS) \cite{intelMDS}. ZombieLoad \cite{schwarz2019zombieload} demonstrates more variants of the line fill buffer leakage (MFBDS), whose secret access is triggered by a microcode assist. Fallout \cite{canella2019fallout} leaks the secret stored in the store buffer called Microarchitectural Store Buffer Data Sampling (MSBDS) \cite{intelMDS}.

A vulerability similar to MDS is the TSX Asynchronous Abort (TAA) \cite{TAA} in Intel processors. If the Intel TSX atomic execution is aborted, uncompleted loads in the transaction may also read a secret from the microarchitectural buffers exploited by MDS and leak it through a covert channel.

The MDS and TAA techniques give rise to more attacks. Uncacheable memory accesses \cite{van2019ridl, intelMDS} can bring data into the buffers mentioned above, which can be accessed using MDS or TAA technqiues and cause the Microarchitectural Data Sampling Uncacheable Memory (MDSUM) attack. The Vector register sampling (VRS) vulnerability \cite{VRS} allows part of the previously accessed vector register values to be sent to the store buffer and get leaked by an MSBDS-type attacker. The CacheOut \cite{cacheout} or L1D eviction sampling (L1DES) vulnerability \cite{intelL1DES} shows that the modified data recently evicted from the L1 data cache can be kept in the line fill buffer, which gives an MFBDS-type attacker the chance to read and leak it. In the CrossTalk \cite{crosstalk} or special register buffer data sampling (SRBDS) attack \cite{intelSRBDS}, the secret value read from certain special registers can be stored in shared buffers and later propagated to the line fill buffer. The secret can be leaked to an MFBDS-type attacker who can even be from a different core. We refer to all the above MDS-related attacks as MDS attacks in \gyreffig{img_attacks_characterization}.

The other type of microarchitectural buffer related attack, i.e., the load value injection (LVI) attacks \cite{vanbulck2020lvi}, explore injecting values to the victim domain to trigger speculation. The attacker first places his malicious data in the microarchitectural buffers and lets the victim access the malicious value through the MDS vulnerabilities. If the malicious value is used by the victim as an address to read a secret or a jump address to an \opaccess-\opuse-\opsend~gadget, the secret can be leaked.

\subsection{Covert Channels for Send Operation}

\begin{figure}[t]
    \centering
    \includegraphics[width=0.8\linewidth]{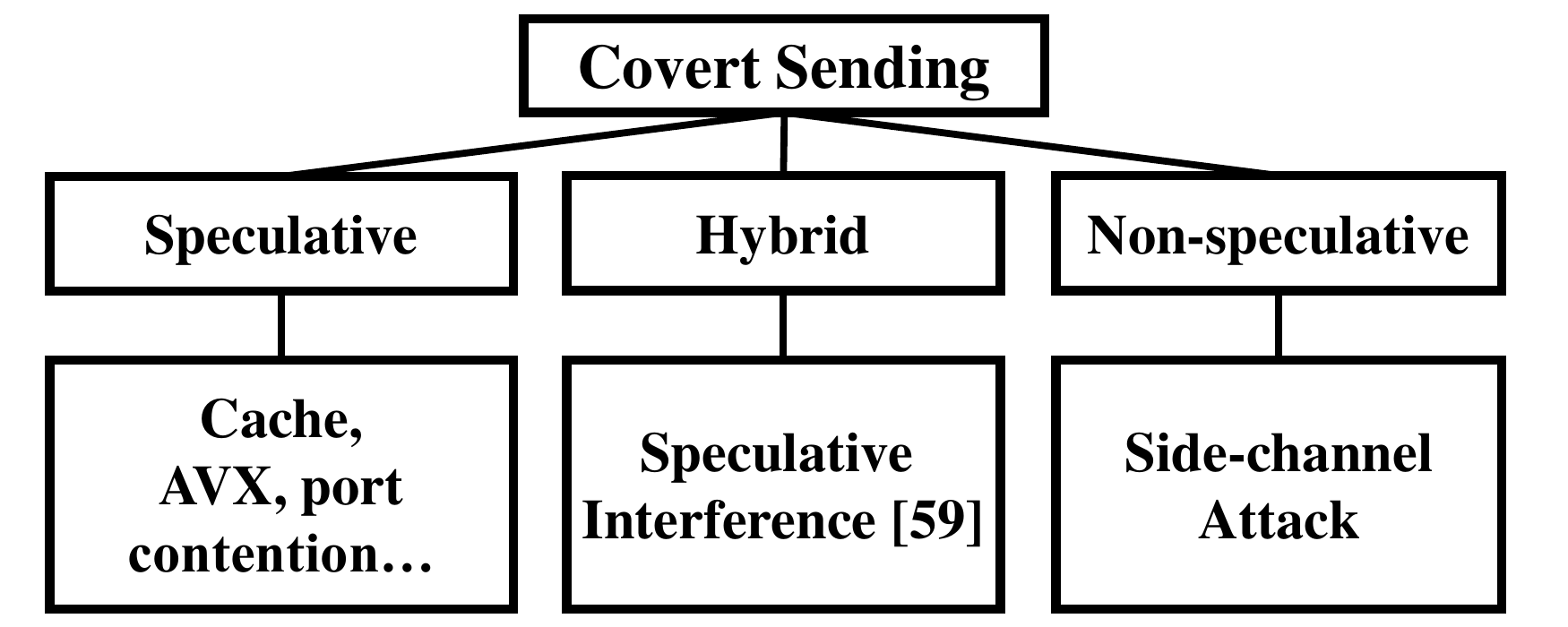}
    \caption{\small{Different ways to leak a secret through a \opsend~operation. The speculative interference attack \cite{speculativeinterference} achieves the final covert \opsend~through a non-speculative instruction.}}
    \label{img_covert_send_taxonomy}
\end{figure}

Microarchitectural covert channels are used to transmit the secret that has been illegally accessed. In \gyreffig{img_covert_send_taxonomy}, we show three types of covert \opsend~operations. These are through speculative instructions, both speculative and non-speculative instructions (hybrid), and only non-speculative instructions. Most of the existing speculative execution attacks are in the first category, executing a speculative \opsend~operation to cause a secret-dependent state change in the covert channel that can be recovered later by the receiver. The cache covert channel is the most commonly used channel. Examples of other covert channels include the execution time of AVX instructions \cite{schwarz2019netspectre}, port contention \cite{bhattacharyya2019smotherspectre} and the cache way predictor \cite{l1dCollPro}.

The recently discovered speculative interference attack \cite{speculativeinterference} leaks the secret through non-speculative instructions by changing the timing of non-speculative instructions with the speculatively executed instructions. In \gyreffig{img_covert_send_taxonomy}, we characterize it as doing a hybrid two-step covert sending. In the first step, the speculative execution causes a secret-dependent hardware unit usage, affecting the timing of non-speculative instructions. In the second step, the timing information of non-speculative instructions can leak the secret. Examples include using the speculative 1) miss status handling register (MSHR) or 2) execution unit contention (first step) to change the timing of a non-speculative load (second step). 
Essentially, the two examples exploit two different covert channels in the first step, rather than the commonly used flush-reload cache channel.

If the \opsend~operation is purely non-speculative as shown in the last case of \gyreffig{img_covert_send_taxonomy}, the attack becomes a side-channel attack, especially when both \opaccess~and \opsend~operations are also non-speculative. This means the program has side-channel vulnerability that allows the secret access and the operation causing a secret-dependent microarchitectural state change, which is beyond the scope of speculative execution attacks.

\bheading{Takeaway from attack analysis.} 
The important observation we make is that the critical attack steps in \gyrefsec{sec_attack_char} hold for all speculative execution attacks, not just for the Spectre v1 attack.
Moreover, \textbf{any valid combination of delayed authorization, speculative secret access and a covert channel can form a new attack variant.} Based on this characterization of speculative attacks, we propose four defense strategies that prevent these speculative execution attacks from succeeding.

\begin{figure*}[t]
    \centering
    \includegraphics[width=\linewidth]{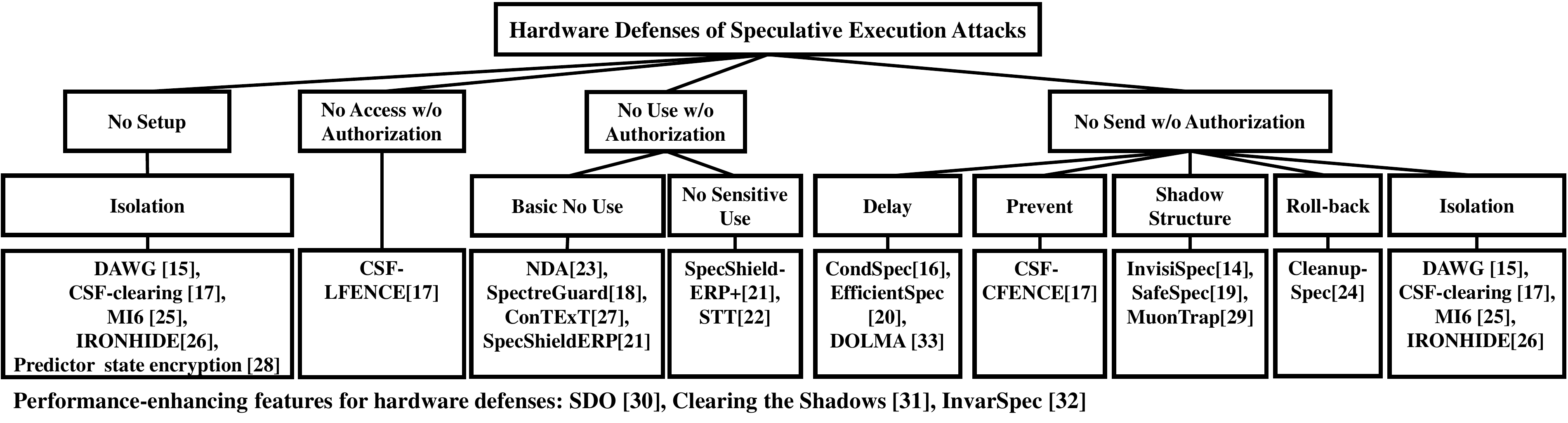}
    \caption{\small{Taxonomy of hardware defenses. The second row shows the 4 defense strategies. The third row shows the child defense categories under each strategy. The fourth row shows the proposed hardware defenses belonging to each defense category.}}
    \label{img_taxonomy}
\end{figure*}

\section{Defense Strategies} \label{sec_taxonomy}

We propose a taxonomy of defenses depending on the attack step prevented, shown in \gyreffig{img_taxonomy}. We identify four defense strategies, each based on a security policy:


\begin{itemize}
    \item \defsetup~(\gyrefsec{sec_no_setup}): \emph{Setup} is prevented so that either the malicious speculative execution cannot start or the covert channel state cannot be initialized.
    \item \defaccess~(\gyrefsec{sec_no_access}): \emph{Access} cannot execute before the authorization is completed.
    \item \defuse~(\gyrefsec{sec_no_use}): \emph{Access} can execute but \emph{Use} of a secret is blocked before the authorization is completed.
    \item \defsend~(\gyrefsec{sec_no_trace}): Both \emph{Access} and \emph{Use} can execute but no secret can be sent, before the secret access is authorized.
\end{itemize}

The insight about \defaccess~is that while \opauth~and \opaccess~may not have any data dependencies, they have a \emph{security dependency} \cite{AttackSummaryPALMSHPCA} since an access should not be allowed until it is authorized. Hence the \defaccess~security policy prevents the security breach. Given that \opaccess, \opuse~and \opsend~are a chain of 3 data-dependent instructions, \defuse~and \defsend~defense strategies can be understood as enforcing the protection at a later stage to try to reduce the performance overhead.

We will describe representative defense proposals for each of these defense strategies. 

\subsection{No Setup}
\label{sec_no_setup}

There are two ways to prevent the \opsetup~step. A defense can prevent either the preparation of the covert channel state or the trigger for speculative execution. Both can be achieved with an isolation-based method shown in \gyreffig{img_taxonomy}.

The isolation method requires partitioning of otherwise shared hardware resources or flushing of a hardware resource if it is time-multiplexed. DAWG \cite{dawg} partitions the cache lines using the domain\_id's and guarantees no interference through the cache replacement state. Context-sensitive fencing \cite{csf} implements a new micro-op to flush the branch target buffers (BTB) or return stack buffer (RSB) state when entering a different protection domain. MI6 \cite{mi6} partitions the shared DRAM and last-level cache (LLC) resources between trusted enclaves and untrusted software and enables clearing any per-core states such as branch predictors, L1 caches and TLBs, with a new instruction. IRONHIDE \cite{ironhide} implements a similar partitioning of LLC and memory resources and also a core-level partitioning by reserving certain cores for a security-critical program to reduce the cost of clearing per-core states.

Encryption can be applied to hardware states to implement an obfuscation-based isolation defense. Predictor state encryption \cite{SamsungExynosCPU} encrypts the BTB or RAS state with a context-specific secret when storing a new target address and decrypts it for usage. This prevents the attacker in another process from injecting malicious jump/return targets, without requiring the clearing of microarchitectural states. Such context-specific encryption can also be considered a form of isolation.

However, note that these \defsetup~defenses usually require that the victim and the attacker come from different security domains, as the isolation-based method uses the domain information to allocate resources and enforce access control and the encryption-based method uses the same key for a certain domain. The same-domain attack, e.g., NetSpectre \cite{schwarz2019netspectre}, cannot be mitigated with these techniques.

\begin{figure}[t]
    \centering
    \includegraphics[width=\linewidth]{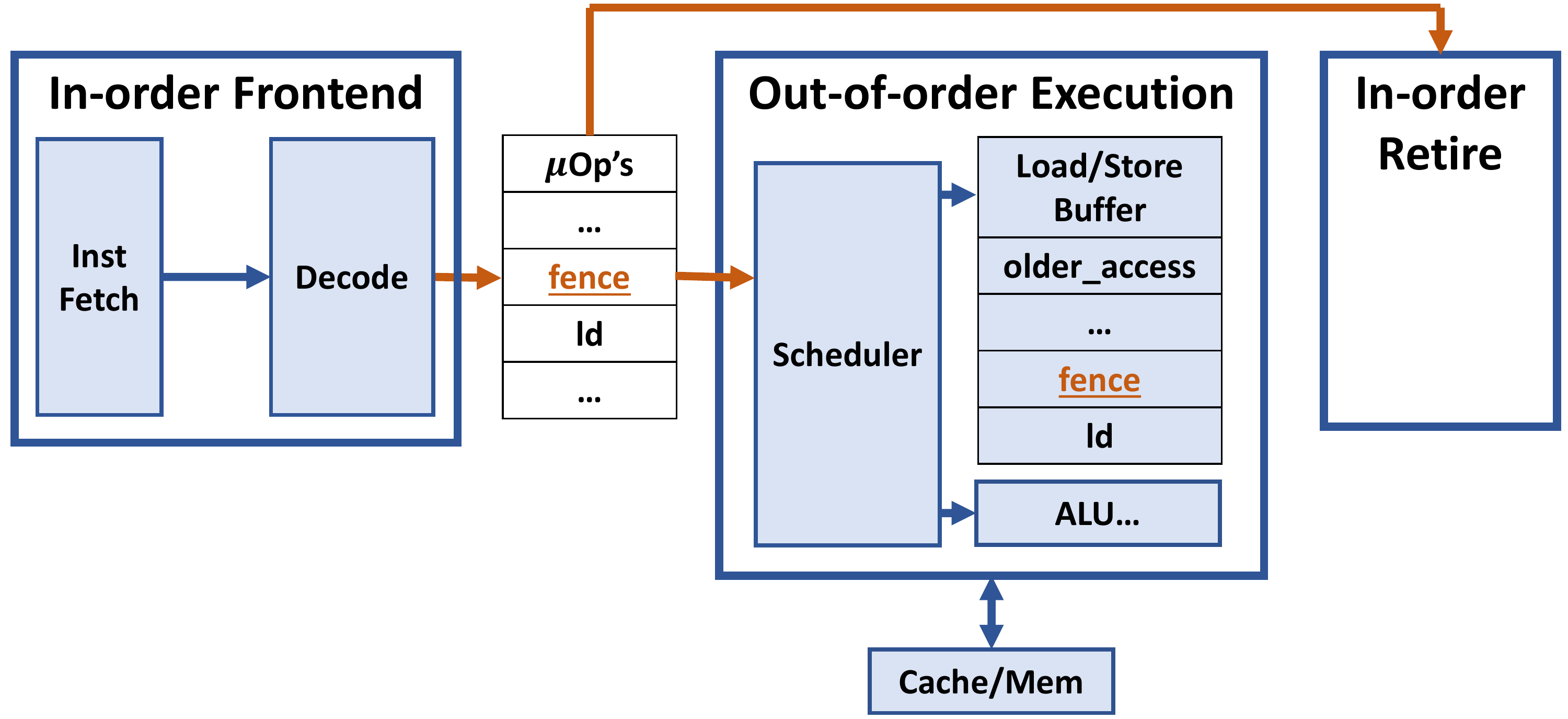}
    \caption{\small{Inserting fences to stall the speculative execution of loads.}}
    \label{img_csf}
\end{figure}

\subsection{No Access Without Authorization}
\label{sec_no_access}

To prevent a security breach, we should prevent the secret \emph{Access} before the authorization is completed. Software solutions can insert memory barriers such as the \emph{lfence} in the x86 ISA to defeat speculative attacks, but they require re-compilation or post-processing of the binary \cite{intel:lfence}. Also, significant performance overhead is incurred with these software fences. A hardware defense can also prevent the secret access by automatically inserting a fence micro-op. Hardware-inserted fences have the advantage of non-intrusive protection and much lower overhead.

The Context-Sensitive Fencing (CSF) defense proposed in  \cite{csf} is shown in \gyreffig{img_csf}. It uses customizable decoding from software instructions to hardware micro-operations to insert hardware fences after a conditional branch instruction before a subsequent load instruction. To defeat the Spectre v1 attack, CSF-LFENCE can place a fence between these two instructions. As no secret data is accessed in the first place, the \defaccess~defense provides strong protection that is independent of the type of covert channel used to exfiltrate the data.

\subsection{No Use without Authorization}
\label{sec_no_use}

Hardware defenses can allow the secret access but prevent its usage in subsequent execution. This improves performance but still blocks the \emph{Use} step in a speculative attack. We call it the \defuse~defense strategy.

\begin{figure}[t]
    \centering
    \includegraphics[width=\linewidth]{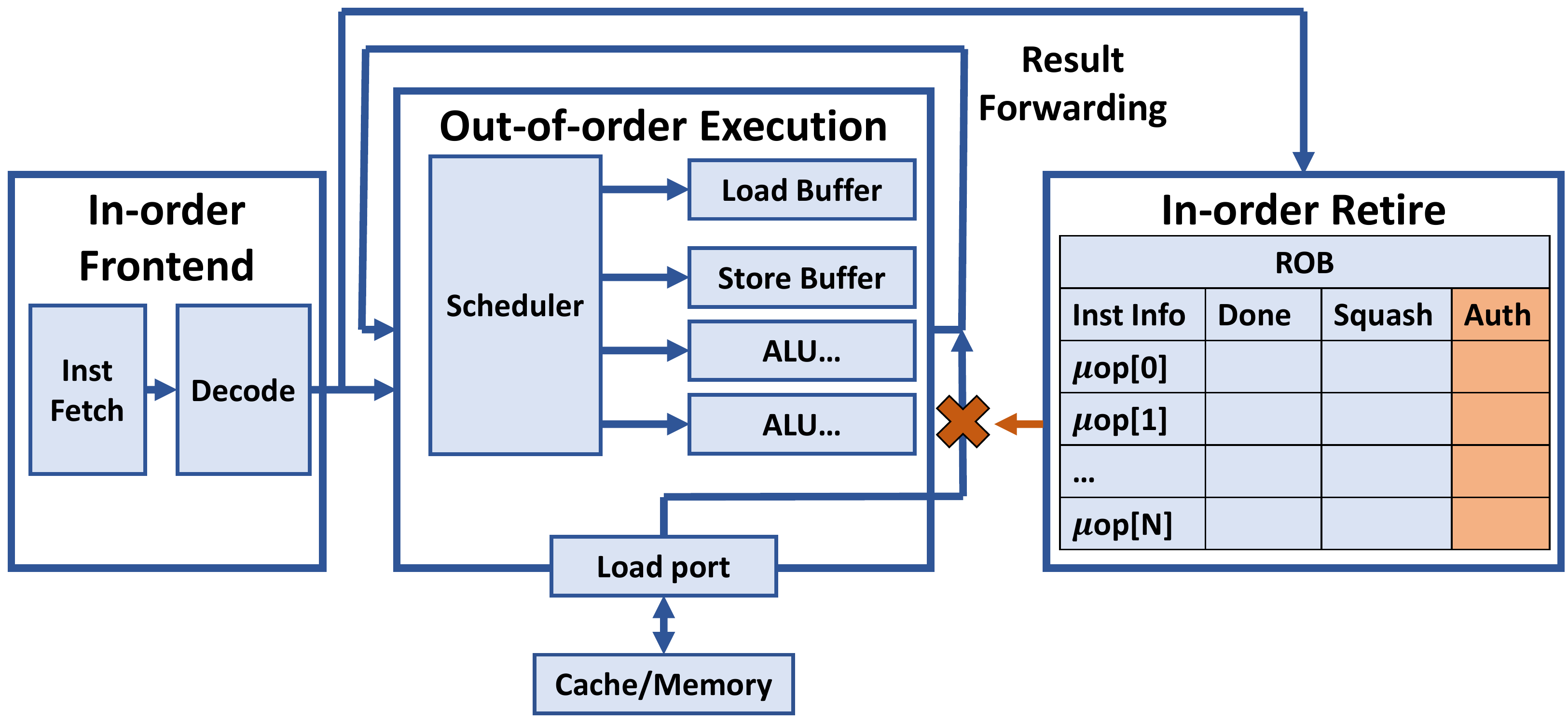}
    \caption{\small{Hardware modification to support \defuse.}}
    \label{img_basicnouse}
\end{figure}

This strategy requires modifying the feed-forward logic which forwards the result of a producer instruction to dependent instructions so that forwarding is allowed to later operations only when the producer instruction is
completed \textit{and} authorized. This can be achieved when both the \textit{Done} and the new \textit{Auth} bits are set in the ROB in \gyreffig{img_basicnouse}.

There are two subclasses of defenses in this category. The ``Basic No Use'' defenses simply prevent the data forwarding to any dependent instructions. The ``No Sensitive Use'' defenses improve the performance by only preventing the data forwarding to sensitive instruction types such as memory load instructions, which can be used to send cache covert channel signals, or for other known covert channels.

\bheading{Basic no use.} An example of the ``Basic No Use'' defense strategy is the NDA  (Non-speculative Data Access) defense proposal \cite{nda}.   This has many variants, based on which authorization checks and \emph{Access} operations are considered. NDA-Permissive checks the resolution of conditional branch conditions and indirect branch addresses (first 2 columns in \gyreffig{img_attacks_characterization}).
 NDA-Permissive-BR (Bypass Restriction) checks these and also checks memory address disambiguation (the third column in \gyreffig{img_attacks_characterization}).
These two NDA-Permissive variants protect accesses from the cache and memory and from special registers like control registers.

There are also two NDA-Strict variants: NDA-Strict and NDA-Strict-BR.  These are like their NDA-Permissive counterparts, except that they also prevent accesses of secrets that are already in the general-purpose registers.

The NDA-Load variant further adds hardware to prevent the data forwarding from an \emph{Access} operation until the instruction is retired, i.e., the instruction is at the head of the ROB queue and has its Authorization completed. 
This covers the first 5 columns in \gyreffig{img_attacks_characterization}. Since NDA was proposed before the last two columns in \gyreffig{img_attacks_characterization}, it is not known if it covers the attacks that do illegal forwarding from microarchitectural buffers. 
NDA-Full is the most secure variant, combining NDA-Strict-BR 
with NDA-Load.

SpectreGuard~\cite{spectreguard} is another example of a ``Basic No Use'' defense. While it only discusses Spectre v1, its key contribution is providing the Linux OS interface to identify sensitive memory pages and mark these as non-speculative. Only data accessed from sensitive pages will not be forwarded during speculative execution, reducing the performance overhead. ConTExT~\cite{context} implements similar software support to mark secret data, which should not be used in speculative execution, as non-transient. In addition, ConTExT allows taint propagation in the processor to also taint the values derived from non-transient values. These tainted values cannot be used in speculative execution that happens in the future.

SpecShield~\cite{specshield} also implements a ``Basic No Use'' defense.  It protects any secret in the memory which can be read by load operations. SpecShieldERP prevents data forwarding until the authorization of control flow, memory disambiguation and memory-related permission checking is completed and no violation is found.

\bheading{No sensitive use.} Another variant in \cite{specshield}, SpecshieldERP+, implements a ``No Sensitive Use'' defense policy by considering the same authorization of control-flow, memory disambiguation authorization and memory permission checking as SpecshieldERP, but only preventing the data forwarding to sensitive instructions like loads and branches. 

Speculative Taint Tracking (STT) \cite{stt} is another example of the ``No Sensitive Use'' policy. STT further considers the covert channels due to implicit information flows and marks loads, branches, stores and data-dependent arithmetic instructions as being sensitive. To improve the performance, STT implements an efficient taint tracking
mechanism to untaint authorized operations.
STT has two variants, STT-Spectre and STT-Future. STT-Spectre considers only the authorization of control flow while STT-future tries to include potential future speculative attacks by deeming a load operation safe only when it reaches the head of the ROB or cannot be squashed.

\subsection{No Send without Authorization}
\label{sec_no_trace}

The \defsend~defenses prevent sending a signal on a covert channel so that the secret cannot be recovered by the attacker, who is the receiver of the covert channel. This signal is sent by changing the microarchitectural state. The defenses under this strategy are usually specific to one or multiple covert channels. Below, we describe five ways to achieve this goal. Although related defense proposals have considered different sets of covert channels, the cache covert channel is the main target that is addressed by all defenses. Hence, we consider specifically the memory load instructions, which change the cache state, to explain these covert channels.

\bheading{Delay state change.} The processor can delay the execution of a load when it needs to modify the cache state.
An example is the Conditional Speculation (CondSpec) defense \cite{condspec},
where an unauthorized memory load that hits in the cache can read the data and complete its execution. However,
a load that has a cache miss is held up to be re-issued later.

The Efficient Invisible Speculative execution (EfficientSpec)  defense \cite{efficientspec} also implements this \textbf{``delay on miss''} mechanism while adding a value predictor to provide a predicted value upon a cache miss.  This  is compared with the real value after the authorization is completed.

The DOLMA defense \cite{dolma} addresses a broader scope of covert channels including not only data caches but also TLBs, instruction caches and hardware predictor state covert channels. It delays both explicit state changes and the changes caused by implicit secret-dependent execution flow and by resource contention. DOLMA considers stores as well as loads as the \opsend~operation.


\bheading{Prevent state change.} The hardware can allow a speculative load to read the data but prevent the cache state change by making the load uncacheable.

Context-sensitive fencing  \cite{csf}, with some variants implementing \defaccess~(Section \ref{sec_no_access}),
also provides a new type of fence, CFENCE, to implement \defsend. A load can execute before a previous CFENCE but it will be converted to a non-cacheable load when it causes a cache miss. This allows the data to be read while preventing the cache state change. The defense variant placing a CFENCE before every load is denoted by ``CSF-CFENCE'' in \gyreffig{img_taxonomy}.

\bheading{Store speculative state in shadow structures.} Visible cache state can be changed only on a successful authorization, by adding a shadow structure to hold the speculatively accessed cache lines.

InvisiSpec \cite{invisispec} prevents the modification of the cache state, including the cache coherence state in the multiprocessor system, by extending the processor with a speculative buffer to store the speculatively accessed data. If the authorization is completed and verified, each speculative load will issue a second access to the same address and cause safe cache state change. If the authorization is completed but rejected, the load is squashed, and no modification is made to the cache state. One InvisiSpec variant, InvisiSpec-Spectre, deems a load unauthorized until all the control-flow predictions are verified. The other variant, InvisiSpec-Futuristic, deems a load unauthorized until it reaches the head of the reorder buffer (ROB) or it cannot be squashed.


The SafeSpec defense \cite{safespec} implements a similar shadow buffer to prevent the modification of both cache and translation lookaside buffer (TLB) states. The cache coherence state is not protected by SafeSpec.

MuonTrap \cite{muontrap} adds the filter caches as the shadow buffers for I-cache, D-cache and TLB. The speculatively accessed entries are only stored in these and get cleared upon security domain switches. A key difference from previous work is that MuonTrap allows non-sensitive modification to the cache coherence state. In a MESI protocol, a speculative access can only be fetched in shared state and any sensitive action changing another cache line from M or E state to S or I state is delayed until it is authorized.

\bheading{Restore state change (Roll-back).} The hardware can allow the cache state change during speculative execution but restore the old cache state if the authorization fails.

CleanupSpec \cite{cleanupspec} prevents a speculative execution attack from modifying the cache state by restoring the cache state when the speculation is found to be wrong. Before the authorization is completed, CleanupSpec allows bringing new cache lines into the cache during speculative execution, but extends each memory request with its side-effect fields to track which cache line is fetched into the cache and which cache line is evicted from the L1 data cache, due to this unauthorized request. If a memory request needs to be squashed, a request is sent to invalidate any new cache line fetched during speculative execution, and bring back any cache line evicted speculatively from the L1 data cache. The L2 and last-level caches in CleanupSpec implement address encryption \cite{ceaser} to prevent eviction-based information leakage.

\bheading{Isolation of states between security domains.} Assuming that the sender and the receiver are from different security domains, some isolation-based defenses that prevent \opsetup~can also prevent the attacker from receiving the covert signaling. For example, the clearing of the branch predictor state can prevent mistraining in the \opsetup~phase and also prevent the leakage through covert sending \cite{branchscope}. Hence, a defense can prevent two steps as a \defsetup~defense and a \defsend~defense.

\begin{table}[t]
\resizebox{\columnwidth}{!}{
\begin{tabular}{|l|l|l|l|l|l|}
\hline
\multirow{2}{*}{\textbf{Feature}}                                                                      & \multirow{2}{*}{\textbf{Enhanced Defense}}                                                                               & \multirow{2}{*}{\textbf{Category}} & \multirow{2}{*}{\textbf{Benchmark}} & \multicolumn{2}{c|}{\textbf{Overhead}}                                                             \\ \cline{5-6} 
                                                                                                       &                                                                                                                          &                                    &                                     & \textbf{Before}                                  & \textbf{After}                                  \\ \hline
SDO \cite{sdo}                                                                        & STT \cite{stt}                                                                                          & No Use                             & SPEC2017                            & About 22\%                                       & 10.05\%                                         \\ \hline
\begin{tabular}[c]{@{}l@{}}ClearShadow\\ \cite{clearshadow}\end{tabular}              & \begin{tabular}[c]{@{}l@{}}Delay on miss\\ \cite{efficientspec}\end{tabular}                            & No Send                            & SPEC2006                            & \multicolumn{2}{l|}{\begin{tabular}[c]{@{}l@{}}9\% faster than basic\\ delay-on-miss\end{tabular}} \\ \hline
\multirow{6}{*}{\begin{tabular}[c]{@{}l@{}}InvarSpec\\ \cite{invarspec}\end{tabular}} & \multirow{2}{*}{fence \cite{invisispec}}                                                                & \multirow{2}{*}{No Access}         & SPEC2006                            & 199.3\%                                          & 101.9\%                                         \\ \cline{4-6} 
                                                                                                       &                                                                                                                          &                                    & SPEC2017                            & 195.3\%                                          & 108.2\%                                         \\ \cline{2-6} 
                                                                                                       & \multirow{2}{*}{\begin{tabular}[c]{@{}l@{}}Delay on miss\\ \cite{condspec, efficientspec}\end{tabular}} & \multirow{2}{*}{No Send}           & SPEC2006                            & 46.1\%                                           & 22.3\%                                          \\ \cline{4-6} 
                                                                                                       &                                                                                                                          &                                    & SPEC2017                            & 39.5\%                                           & 24.4\%                                          \\ \cline{2-6} 
                                                                                                       & \multirow{2}{*}{InvisiSpec \cite{invisispec}}                                                           & \multirow{2}{*}{No Send}           & SPEC2006                            & 18.0\%                                           & 9.6\%                                           \\ \cline{4-6} 
                                                                                                       &                                                                                                                          &                                    & SPEC2017                            & 15.4\%                                           & 10.9\%                                          \\ \hline
\end{tabular}
}
\caption{The improvement in performance overhead by applying SDO, ClearShadow and InvarSpec to existing defenses.}
\label{tbl_performance_optim}
\end{table}

\subsection{Reducing Overhead of Defenses} \label{sec_performance_optim}

Techniques have been proposed to reduce the performance overhead of defenses described earlier. \gyreftbl{tbl_performance_optim} shows the performance improvements they achieve.

Speculative Data-Oblivious Execution (SDO) \cite{sdo} allows an instruction, which may depend on a secret, to execute. For instance, a speculative load can access certain cache levels without making any state changes and the performance is improved if the data is found. SDO can be integrated with STT \cite{stt}.

Clearing the Shadows (ClearShadow) \cite{clearshadow} improves the performance by accelerating the computation of branch conditions and memory addresses so that \opauth~can finish earlier. ClearShadow moves the instructions that \opauth~depends on to the front to shorten or remove the speculation window. ClearShadow has been used to improve a ``delay-on-miss'' defense \cite{efficientspec}.

InvarSpec \cite{invarspec} allows some sensitive instructions to execute earlier without protection. InvarSpec software identifies the safe set (SS) of an instruction $I$ which contains instructions that are older than $I$ but do not affect $I$'s input and execution. InvarSpec hardware extension reads the SS and allows $I$ to be issued even if some SS instructions are not resolved. InvarSpec can be applied to the fence-based defense, the delay-on-miss defense and the InvisiSpec defense as we show in \gyreftbl{tbl_performance_optim}.

\subsection{Software-hardware Co-design}
\label{sec_codesign}

Some hardware defenses require software support. One way is changing the application software as described above for ClearShadow \cite{clearshadow} and InvarSpec \cite{invarspec}. Another way is modifying the system software. DAWG \cite{dawg} needs the system software to assign a proper domain ID to the protected program so that the domain ID is not shared with any potential attackers. Context-sensitive fencing \cite{csf} has a set of model-specific registers (MSRs) to specify the fence type and the insertion strategy. SpectreGuard \cite{spectreguard} and ConTExT \cite{context} enable marking secret data as non-transient by using a bit in the page table entry, which requires both compiler and OS software modifications. 


\begin{table}[t]
\resizebox{\columnwidth}{!}{
\begin{tabular}{|l|l|l|l|}
\hline
\textbf{Strategy}                                                              & \textbf{Defense}                                    & Platform                                                                                            & \textbf{\begin{tabular}[c]{@{}l@{}}Performance\\ Overhead (\%)\end{tabular}}        \\ \hline
\multirow{3}{*}{\begin{tabular}[c]{@{}l@{}}No Setup \&\\ No Send\end{tabular}} & DAWG \cite{dawg}                   & Zsim \cite{zsim}                                                                   & 0 $\sim$15                                                                          \\ \cline{2-4} 
                                                                               & MI6 \cite{mi6}                     & RiscyOO \cite{riscyoo}                                                             & 16.4                                                                                \\ \cline{2-4} 
                                                                               & IRONHIDE \cite{ironhide}           & \begin{tabular}[c]{@{}l@{}}Tilera Tile-Gx72\\ processor \cite{tilera}\end{tabular} & \begin{tabular}[c]{@{}l@{}}-20\\ (Compared to an \\ SGX-like baseline)\end{tabular} \\ \hline
No Access                                                                      & CSF-LFENCE \cite{csf}              & GEM5 \cite{gem5}                                                                   & 48                                                                                  \\ \hline
\multirow{5}{*}{No Use}                                                        & NDA \cite{nda}                     & GEM5 \cite{gem5}                                                                   & 10.7 $\sim$125                                                                      \\ \cline{2-4} 
                                                                               & SpectreGuard \cite{spectreguard}   & GEM5 \cite{gem5}                                                                   & 8, 20                                                                               \\ \cline{2-4} 
                                                                               & ConTExT \cite{context}             & \begin{tabular}[c]{@{}l@{}}Software approximation\\ on Intel processor\end{tabular}                 & 0.1 $\sim$71.1                                                                      \\ \cline{2-4} 
                                                                               & SpecShieldERP(+) \cite{specshield} & GEM5 \cite{gem5}                                                                   & 10, 21                                                                              \\ \cline{2-4} 
                                                                               & STT \cite{stt}                     & GEM5 \cite{gem5}                                                                   & 8.5, 14.5, 24, 27                                                                   \\ \hline
\multirow{8}{*}{No Send}                                                       & CondSpec \cite{condspec}           & GEM5 \cite{gem5}                                                                   & 6.8, 12.8, 53.6                                                                     \\ \cline{2-4} 
                                                                               & EfficientSpec \cite{efficientspec} & GEM5 \cite{gem5}                                                                   & 11 (IPC loss)                                                                       \\ \cline{2-4} 
                                                                               & DOLMA \cite{dolma}                 & GEM5 \cite{gem5}                                                                   & 10.2 $\sim$42.2                                                                     \\ \cline{2-4} 
                                                                               & CSF-CFENCE \cite{csf}              & GEM5 \cite{gem5}                                                                   & 7.7, 21                                                                             \\ \cline{2-4} 
                                                                               & InvisiSpec \cite{invisispec, invisispec:correction}       & GEM5 \cite{gem5}                                                                   & 5, 17                                                                               \\ \cline{2-4} 
                                                                               & SafeSpec \cite{safespec}           & MARSSx86 \cite{marssx86}                                                           & -3                                                                                  \\ \cline{2-4} 
                                                                               & MuonTrap \cite{muontrap}           & GEM5 \cite{gem5}                                                                   & -5, 4                                                                               \\ \cline{2-4} 
                                                                               & CleanupSpec \cite{cleanupspec}     & GEM5 \cite{gem5}                                                                   & 5.1                                                                                 \\ \hline
\end{tabular}
}
\caption{Performance numbers reported by existing work. The numbers may not be directly comparable as they are measured in different configurations. Numbers separated by commas are for different defense variants or benchmarks.}
\label{img_perf_general}
\end{table}

\section{Understanding Performance Overhead}
\label{sec_performance}

\subsection{Performance Overhead Reported by Defense Papers}
TABLE~\ref{img_perf_general} shows the performance overhead reported by some hardware defenses, listed according to the hardware defense taxonomy we presented in Fig.~\ref{img_taxonomy}. 
The same gem5 cycle-accurate processor simulator \cite{gem5} is used by most of the hardware defense papers. The overhead of isolation-based defenses to prevent cross-domain \opsetup~and \opsend~is mainly due to the clearing of microarchitectural states and the partitioning of hardware resources. CSF-LFENCE \cite{csf} inserts lfence for only kernel loads but already incurs an overhead of 48\%. The \defuse~defenses differ a lot in their performance overhead as they may cover different types of authorization (NDA), protect certain data region (SpectreGuard), be emulated with software (ConTExT), and prevent certain sensitive \opuse's (SpecShield and STT). The \defsend~defenses generally have lower perfromance overhead as they only address certain covert channels, especially the cache covert channel.

\begin{table}[t]
    \centering
    \includegraphics[width=\linewidth]{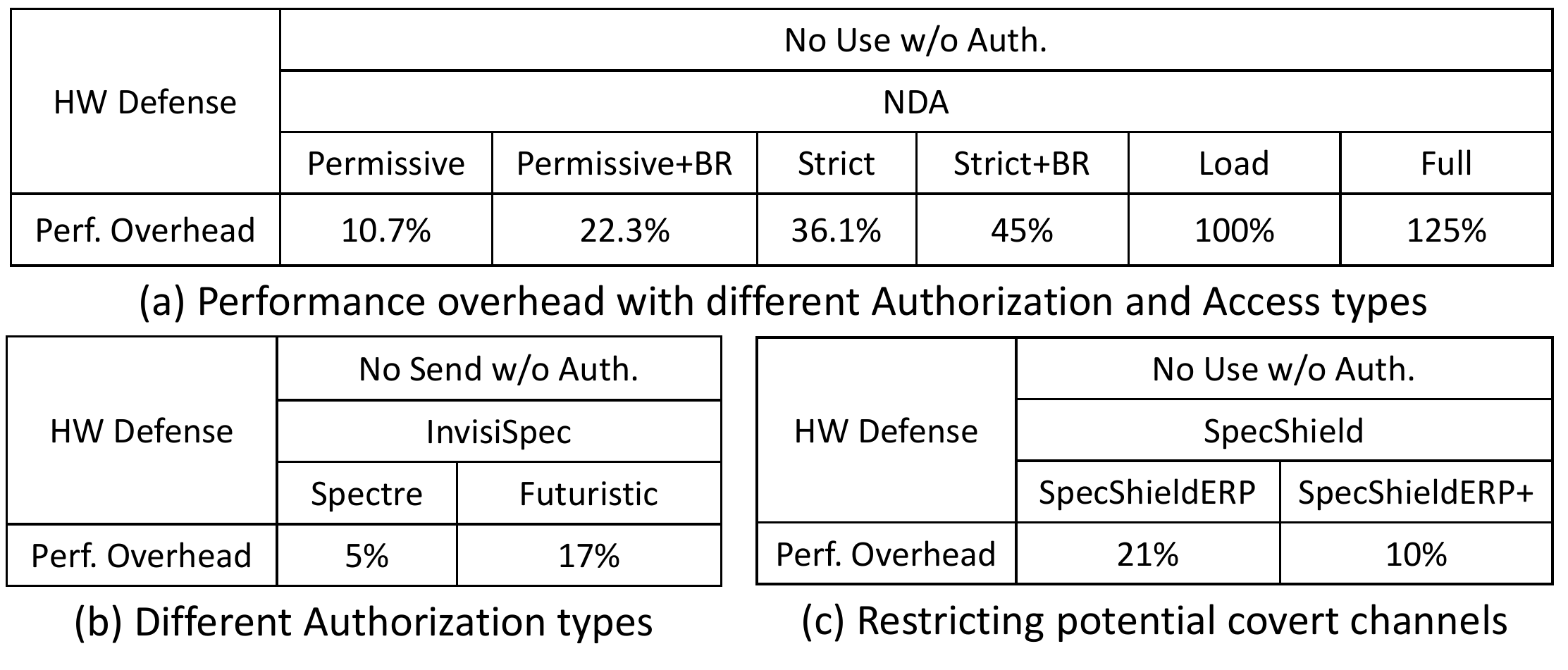}
    \caption{\small Security-performance trade-offs of different variants within the same work.}
    \label{img_perf_within_work}
\end{table}


We illustrate how some of the defenses trade off security and performance. For increased security, more attacks and vulnerabilities can be covered, and more covert channels mitigated, but at increased performance overhead.

\bheading{Increased overhead for covering more attacks.}~\gyreftbl{img_perf_within_work}(a) shows the increase in performance overhead for the NDA \cite{nda} defense variants to prevent more types of attacks. The ``Permissive'' variant considers the control flow authorization only (the first two columns in \gyreffig{img_attacks_characterization}). The ``Permissive+BR (Bypass Restriction)'' variant further considers memory disambiguation authorization (the third column in \gyreffig{img_attacks_characterization}). ``Load'' (NDA-Load) considers the first five columns in \gyreffig{img_attacks_characterization} by not deeming an \opaccess~operation authorized until it is retired. A fair comparison of performance overhead is from ``Permissive'' (10.7\%), to ``Permissive+BR'' (22.3\%), then to ``Load'' (100\%), since they all protect secrets in memory and special registers. 

InvisiSpec \cite{invisispec} provides two variants: InvisiSpec-Spectre
defends against control-flow misprediction based attacks and InvisiSpec-Futuristic tries to defend against future attacks, where any speculative load may pose a threat.
The latter one is more secure but has more performance overhead (17\% vs. InvisiSpec-Spectre's 5\% in \gyreftbl{img_perf_within_work} (b)) \cite{invisispec:correction}. 

\bheading{Access type vs. Performance trade-off.} TABLE~\ref{img_perf_within_work}(a) also shows that as more types of \emph{Access} are considered, the performance overhead increases.
The variant ``Strict+BR'' considers the accesses to general-purpose registers (GPRs) in addition to special registers and memory, which are considered by ``Permissive+BR''.
The overhead increases from 22.3\% (Permissive+BR) to 45\% (Strict+BR) due to the GPR access consideration.

\bheading{Mitigated covert channels vs. Performance trade-off.} In \gyreftbl{img_perf_within_work} (c) which compares two SpecShield \cite{specshield} variants, SpecShieldERP disallows the forwarding from a speculative load to all instructions while SpecShieldERP+ only disallows the forwarding to sensitive loads and branch instructions, which may be covert \opsend's.  
SpecShieldERP+ relaxes some
of the security guarantees 
to reduce the
performance impact of SpecShieldERP from 21\% to 10\%.

\subsection{Security-Performance Tradeoffs Considering Our Defense Strategies} \label{sec:perf-analysis}

\begin{figure}[t]
    \centering
    \includegraphics[width=\linewidth]{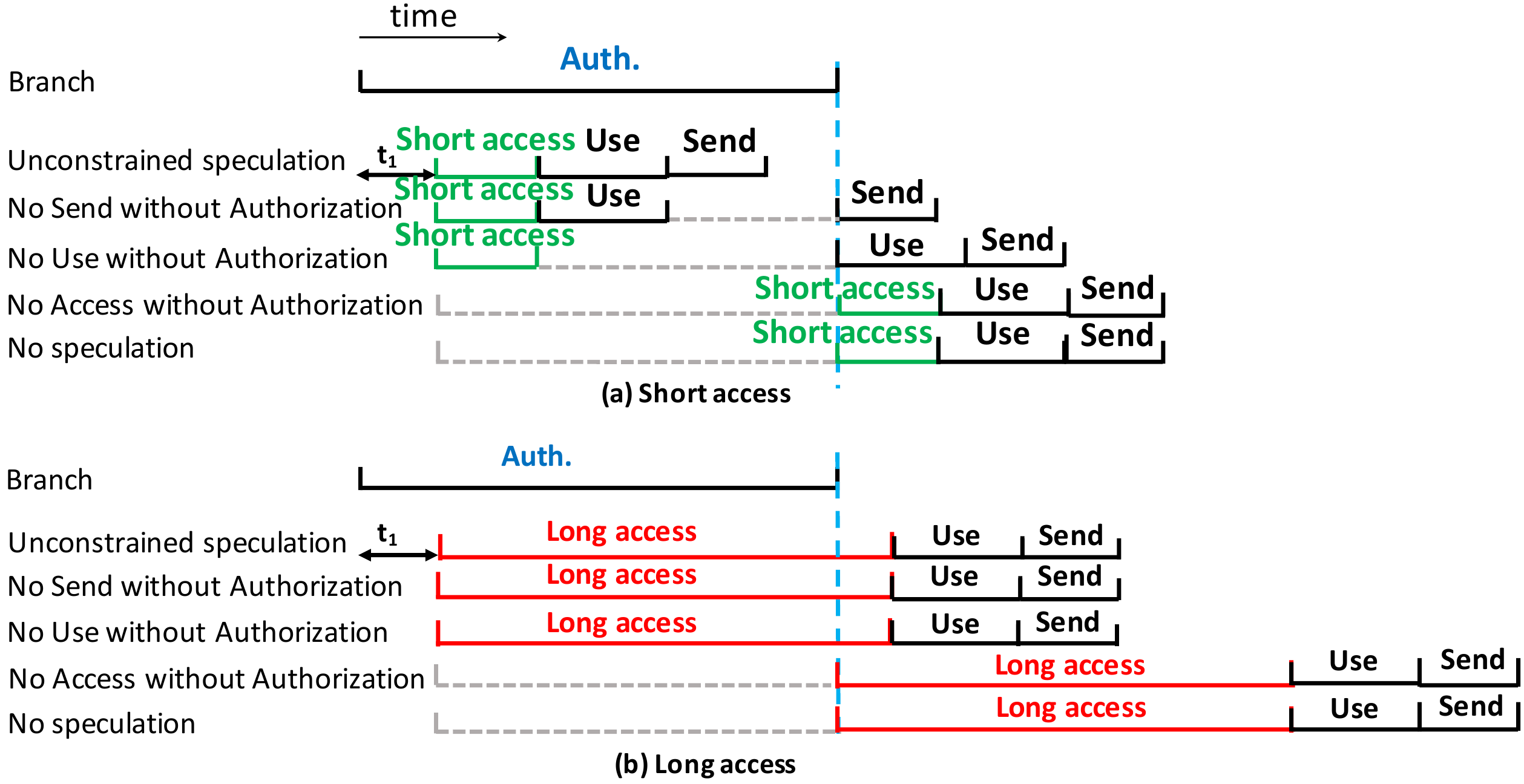}
    \caption{Performance analysis of different speculative execution defense strategies on a fast access (a) and a slow access (b).}
    \label{fig:perf-analysis:defense-timeline}
\end{figure}

We now consider the theoretical performance overhead reductions that might be expected, as we relax the security policy from \defaccess~to \defuse~to \defsend. These correspond to the three main categories in our defense strategies in \gyreffig{img_taxonomy}. Since speculative attacks are rare, our goal is to compare the impact of these defense strategies on normal (benign) speculative execution. We consider an example benign program containing a branch instruction, a first load instruction, an arithmetic instruction and a second load instruction, that is data dependent on the arithmetic instruction which is data dependent on the first load instruction. While there may be an arbitrary number of instructions between these 3 instructions, we show them as sequential (in \gyreffig{fig:perf-analysis:defense-timeline}), to simplify the discussion.
To help correlate this code with a speculative execution attack, e.g., Spectre v1 in \gyreffig{img_code_spectre}, these 4 instructions correspond to the \opauth, \opaccess, \opuse~and \opsend~operations.

We illustrate the timelines of the \opauth, \opaccess, \opuse~and \opsend~operations in Fig. \ref{fig:perf-analysis:defense-timeline}. We consider two scenarios: a fast secret access (Fig. \ref{fig:perf-analysis:defense-timeline}(a)), e.g., the first load has a cache hit, and a slow secret access (Fig. \ref{fig:perf-analysis:defense-timeline}(b)), e.g., the first load has a cache miss. In each scenario, we present 1) an insecure OoO processor allowing any speculation (\fullspec); 2) a \defsend~defense; 3) a \defuse~defense; 4) a \defaccess~defense; and 5) a processor disabling speculation (\defnospec).

The \opaccess, \opuse~and \opsend~are a chain of three data-dependent instructions. Therefore the \opaccess, \opuse~and \opsend~operations cannot run together. In Fig. \ref{fig:perf-analysis:defense-timeline}(a)), the \defsend, the \defuse, and the \defaccess~strategies delay the \opsend, \opuse~and \opaccess~operations, respectively, till after the Authorization is resolved.  Hence they have increasing performance overhead, showing
the intrinsic performance overhead in these defense strategies. 
The slowest but most rigorous security policy, \defaccess, is as slow as \defnospec, if the first load instruction (\opaccess) immediately follows the branch instruction, and there are no other non-dependent instructions that can be executed in the speculation window.

In Fig. \ref{fig:perf-analysis:defense-timeline}(b), the slow \opaccess~(cache miss) may not be fully covered by the delay in the \opauth~operation. The \defnospec~defense still causes the longest delay, same as the \defaccess~defense strategy. The \defuse~and the \defsend~defense strategies introduce shorter delay, and can achieve the fastest performance like the \fullspec~case.

\bheading{Takeaway:} In general, the \defnospec~and \fullspec~cases give the upper and lower bounds, respectively, for the total execution time. The overheads of the three defense strategies decrease from the strict security policy of \defaccess, to the more relaxed but still secure \defuse, to the \defsend~strategies, with not much difference in overhead between the last two strategies.
\section{Problems for Some Defenses}

\bheading{Problem scenarios for isolation-based defenses} The isolation-based defenses can be used to either prevent the \opsetup~step (\gyrefsec{sec_no_setup}) or the covert \opsend~step (\gyrefsec{sec_no_trace}). These defenses can prevent the attack when the victim and the attacker are from different security domains. However, the mis-training can happen in the same domain \cite{AttackSummaryGraz:236214, schwarz2019netspectre}. The sender and the receiver of covert channel communication can also be from the same domain. For instance, in a Meltdown attack demo \cite{MeltdownRepoIAIK} where the secret is in the kernel space, the sender instructions that read the secret and send it out are in a user-level process, which also executes the receiver's code to reveal the secret. These special cases can make the isolation-based defenses ineffective. For instance, the DAWG \cite{dawg} cache uses the domain ID to parition the cache resouces and therefore, it cannot prevent the same-domain attack where the sender and the receiver are in the same process and have the same domain ID. Other \defsend defenses may also have to be applied, e.g., defenses following \textit{Delay}, \textit{Prevent}, \textit{Shadow Structure} or \textit{Roll-back} in \gyrefsec{sec_no_trace}.

\bheading{Problem with covert channel blocking defenses.} The issue with the \defsend~defenses is that they only protect against one or a few covert channels. However, they do not restrict how secrets are illegally accessed and can protect against new speculative execution or other attacks - but only for the specific covert channels considered in the defense.

\section{Recommendations and Future Work}
\label{sec_discussion}

\bheading{Recommendation of defense strategies.} From a security perspective, preventing the \opaccess~and \opuse~of a secret is more critical, since all covert channels requiring secret-dependent usage are eliminated. Between these two strategies, the \defuse~has better performance as some long-latency loads can be performed under speculative execution, reducing the performance overhead in benign situations.

\bheading{Mitigating the aggressive forwarding of faulty access.} The faulty access attacks can read data that the current program does not have the permission to access. Some defenses prevent \cite{csf, nda} these attacks by blocking the execution or completion of an \opaccess~operation until it is at the head of the reorder buffer (ROB) so that any exception will be immediately handled. We argue that for the faulty access that violates the permission check of memory or special registers, delaying the forwarding to any dependent instructions until its permission check is finished is enough, i.e., a \defuse~policy. 

For the illegal forwarding from microarchitectural buffers, our suggestion is to disallow the forwarding to any faulting memory accesses, or return a dummy value and disallow its usage. Simply returning a dummy value without preventing its usage is not enough. For instance, returning a dummy value of 0 may cause the leakage of the data at address 0 if the dummy value is speculatively used as an address.

\section{Conclusions}
\label{sec_conclusion}

In this paper, we first show how speculative execution attacks can be classified according to what hardware features are exploited to bypass security checks, that we call
\textit{authorizations}.  We then show a new attack characterization based on the critical attack steps common to all speculative execution attacks, namely, \emph{Setup}, \emph{Authorize}, \emph{Access}, \emph{Use}, \emph{Send} and \oprecv. We observe that the root cause of the attacks succeeding is the bypassing of the \emph{Authorize} step
during speculative execution. This attack characterization enables us to propose the first taxonomy of defense strategies, where each strategy prevents one of the critical attack steps of \emph{Setup}, \emph{Access}, \emph{Use} and \emph{Send}. We show that the 20 defense proposals considered in this paper can be categorized under at least one of these four defense strategies, or as an overhead-reducing feature.  We describe important features in these defenses and some of their key hardware modifications. Security-performance tradeoffs are also discussed for defenses that propose multiple variants. We discuss the scope of these hardware defense strategies and show their 
relative performance overhead. 

Future work can consider new attacks and defenses, using and  adding to our taxonomies of attacks and defenses. New defenses can be proposed to reduce the performance overhead and/or cover more attack types. For fair comparisons, new defenses should compare their performance with those that target the same set of exploited vulnerabilities, secret accesses and covert channels. 


\bheading{Acknowledgements.} This work was supported in part by NSF SaTC \#1814190, SRC Hardware Security \#2844 and a Qualcomm Faculty Award for Prof. Lee. We thank Shuwen Deng and Jakub Szefer for help with initial performance numbers.

\newpage

\bibliographystyle{IEEEtran}
\bibliography{refs}

\end{document}